\documentclass[openacc]{rstransa}%%%%where rstrans is the template name

\DeclareUnicodeCharacter{FB01}{fi}

\def\tdist{D_{\Delta t}}
\def\Dd{D_{\rm d}}
\def\Dds{D_{\rm ds}}
\def\Ds{D_{\rm s}}
\def\zd{z_{\rm d}}

\titlehead{Review}

\begin{document}

%%%% Article title to be placed here
\title{Challenges and Opportunities for time-delay cosmography with multi-messenger gravitational lensing}

\author{%%%% Author details
Simon~Birrer$^{1}$, Graham~P.~Smith$^{2, 3}$, Anowar~J.~Shajib$^{4,5, 6}$, Dan~Ryczanowski$^{7,2}$, and Nikki~Arendse$^{8}$}

%%%%%%%%% Insert author address here
\address{affiliations at the end of the manuscript
}

%%%% Subject entries to be placed here %%%%
\subject{Cosmology, gravitational waves, Hubble constant, gravitational lensing}

%%%% Keyword entries to be placed here %%%%

%%%% Insert corresponding author and its email address}
\corres{Simon Birrer\\
\email{simon.birrer@stonybrook.edu}}

%%%% Abstract text to be placed here %%%%%%%%%%%%
\begin{abstract}
\sloppypar
Strong gravitational lensing of variable sources, such as quasars or supernovae, can be used to constrain cosmological parameters through a technique known as ``time-delay cosmography''. Competitive constraints on the Hubble constant have been achieved with electromagnetic observations of lensed quasars and lensed supernovae. Gravitational wave (GW) astronomy may open up a new channel for time-delay cosmography with GW signal replacing the electromagnetic (EM) one. 
We highlight the similarities of using GW signals to be applied to time-delay cosmography compared to EM signal. We then discuss key differences between GW and EM signals and their resulting advantages and inconveniences from the angle of the current state-of-the-art using quasars and lensed supernovae for time-delay cosmography. We identify the astrometric precision requirement of the images as a key challenge to overcome and highlight the potentially significant impact that near-perfect time-delay measurements of lensed GWs can bring to the table. 
\end{abstract}
%%%%%%%%%%%%%%%%%%%%%%%%%%%

%%%%%%%%%% Insert the texts which can accomdate on firstpage in the tag "fmtext" %%%%%

\begin{fmtext}

\end{fmtext}

\maketitle

\section{Introduction}

Strong gravitational lensing, when the bending of light, or any propagating wave, such as a gravitational wave, leads to more than one appearing image, results in a relative arrival time difference between the arriving wave fronts.
The relative arrival time is directly proportional to the scales in the universe. The method that uses strong gravitational lensing of variable sources, such as quasars or supernovae, to constrain cosmological parameters is known as ``time-delay cosmography''. By measuring the time delay between multiple lensed images of the source, it is possible to constrain the ``time-delay distance'' of the lens system and thus $H_{0}$ \cite{Refsdal:1964}. 

% importance of H0 measurement
The current local expansion rate of the universe, the Hubble constant $H_0$, is a key quantity to determine cosmological distances, the size and age of the universe. At this stage, there are different significantly discrepant measurement of $H_0$. On one side, the local direct measurements of $H_0$ \cite{Riess:2022}, and on the other hand the measurements anchored on early-universe physics and distances and model-extrapolated to current time \cite{Planck:2020}.
Time-delay cosmography provides a completely independent and direct measurement of distances. A precise and accurate measurement of $H_0$ with time-delay cosmography can shed light on whether either new physics is required to provide a consistent interpretation of early- and late-universe measurements, or whether there are unaccounted-for systematics uncertainty in either of the measurements (see e.g.,  \cite{DiValentino:2021, Abdalla:2022} for a review).

The to-date most prominent type of variable background source used for time-delay cosmography are quasars due to their electromagnetic brightness and variability on short timescales (e.g., \cite{Suyu:2010, Wong:2020, Birrer:2020}).
Results using lensed supernovae have also recently shown competitive constraints \cite{Kelly:2023, Pascale:2024}.
The field of time-delay cosmography has matured over the past two decades, and a dedicated effort to perform the necessary measurements and controlling for systematics is underway, with an anticipated $\sim 1\%$ measurement of the Hubble constant with a sample of $\sim 40$ lensed quasars \cite{Birrer:2021}. 

% lensed Ia's from LSST
With LSST, several dozens of lensed supernovae (SNe) are expected to be detectable in the data per year (e.g., \cite{Wojtak:2019, Arendse:2023}) and forecasts including follow-up observations project a similar $\sim 1\%$ precision on $H_0$ at the end of the 10 years LSST survey
\cite{Birrer:2021glSNe, Suyu:2020}.
We also refer to\cite{Suyu:2024} for a recent review on strongly lensed SNe for applications to cosmography and astrophysics.

Gravitational wave (GW) astronomy may open up a new channel for time-delay cosmography, with the GW signal replacing the electromagnetic (EM) signal. 
This channel for time-delay cosmography has been recognised in the literature with a variety of techniques and forecasts explored (e.g., \cite{Liao:2017, Wei:2017, Li:2019, Hou:2020, Jana:2023}).

In this manuscript, we highlight the similarities of using GW signals compared to the current quasars and supernovae to be applied to time-delay cosmography. We then discuss key differences between GW and EM signals and their resulting advantages and inconveniences from the angle of the current state-of-the-art using lensed quasars and lensed supernovae.
In Section~\ref{sec:tdc}, we briefly review the general principles of time-delay cosmography. In Section~\ref{sec:gw}, we present the challenges and opportunities for lensed GW time-delay cosmography. In particular, we will highlight the opportunities with an extremely precise time-delay measurement and the use of small separation strong lenses, and the anticipated challenges with the astrometric precision. 
In Section~\ref{sec:forecast} we perform a more detailed forecast with a combination of time-delay, astrometric and lens model error budgets. We summarise our findings in Section~\ref{sec:summary}.

\section{Time-Delay Cosmography} \label{sec:tdc}
In this section, we summarise the principles behind time-delay cosmography and refer the reader to, e.g., \cite{Birrer:2022} and \cite{Treu:2022} for recent comprehensive reviews focusing on time-delay cosmography applied to EM signals and \cite{Saha:2024} for providing the essentials on the theory of strong gravitational lensing.
Any electromagnetic (EM) or gravitational wave (GW) emitted from a background source experiences a delay in arrival time at the observer because of the gravitational lensing effect.  This time delay is the result of two physical effects: the excess path length of the deflected wave relative to the path it would take in the absence of lensing, and the time dilation arising from the wave passing through the gravitational potential of the lens \cite{Shapiro:1964}.  The excess time delay for a given lensed image (relative to an image that is not lensed) is 
\begin{equation} \label{eq:excess_td}
t(\vec{\theta}, \vec{\beta}) = \frac{\tdist}{c} \phi(\vec{\theta},\vec{\beta}),
\end{equation}
where $\vec{\theta}$ is the image position, $\vec{\beta}$ is the source position, $\phi$ is the ``Fermat potential'' at the image position, $c$ is the speed of light, and $\tdist$ is the time-delay distance.  The Fermat potential, which represents the arrival time surface of wavefronts, is
\begin{equation} \label{eq:fermat}
\phi(\vec{\theta},\vec{\beta}) = \frac{(\vec{\theta} - \vec{\beta})^{2}}{2} - \psi(\vec{\theta}),
\end{equation}
where $\psi$ is the deflection potential that is related to the deflection angle $\vec{\alpha}$ by $\nabla \psi \equiv \vec{\alpha}$. 
Fermat's principle, that wavefronts arrive at extrema of the arrival time surface, $\nabla \phi(\vec{\theta})=0$, results in the lens equation
\begin{equation}
    \vec{\beta} = \vec{\theta} - \vec{\alpha}(\vec{\theta}).
\end{equation}
The time-delay distance is defined as
\begin{equation} \label{eq:ddt}
\tdist \equiv (1+\zd) \frac{\Dd \Ds}{\Dds},
\end{equation}
where $\zd$ is the lens redshift, and $\Dd$, $\Ds$, and $\Dds$ are the angular diameter distances from the observer to the lens, the observer to the source, and the lens to the source, respectively.  
% absolute distance
The angular diameter distances in Equation~\ref{eq:ddt} have absolute units of distance and hence are a measure of the absolute scale of the Universe. The distance measure is inversely proportional to $H_0$, that is, $\tdist \propto H_{0}^{-1}$, and hence $H_0$ is the primary parameter that can be determined directly from measurements of the time-delay distance.

In order to constrain $\tdist$ (and therefore, $H_{0}$) from a multiply-imaged time-variable source, one must measure the arrival time difference between at least one pair of images, and constrain the Fermat potential at the position of the same images from an accurate mass model of the lens.  
In practice, for lensed quasars, this is done by monitoring the lens over a period of time and comparing the light curves of the multiple images to look for common features corresponding to the same brightness fluctuation at the source. For lensed supernovae, immediate follow-up is required, with space telescopes. Time delays can be estimated with time series of photometric filters \cite{Kelly:2023_time_delay, Pierel:2024}, as well as with spectroscopy in the case of a type Ia SN \cite{Chen:2024}. The transient nature and regularity of the SNe allow a measurement of the time delay with fewer, but well-timed observations. The more regular and known the intrinsic brightness fluctuations are, the more precise a time delay can be estimated. In the case of a GW event, the time-variable behaviour is very well understood and hence will allow for a very precise time-delay measurement.

\section{Challenges and opportunities for lensed GW}\label{sec:gw}
All equations presented in Section~\ref{sec:tdc}, in particular the Fermat principle, are valid in wave optics for GWs on scales of galaxies, when the size of the lenses is larger than the wavelength of the GW. In this section, we highlight some key opportunities and challenges of lensed GW to conduct time-delay cosmography measurements.

\subsection{Time-delay measurement}\label{sec:time-delay}
Gravitational wave detectors have a time resolution of order of a millisecond. Therefore, the well-understood GW binary merger waveform allows for a precise arrival time measurement of order millisecond for any arriving image detected with a sufficient signal-to-noise ratio. To put this into perspective, the most precise time-delay measurement with an EM signal is just slightly lower than a day, reflecting the limit in the cadence of the observations.

In effect, GW timing precision translates to $\lesssim1$cm/s/Mpc precision on $H_0$ for arrival-time differences of $\gtrsim1\,\rm hours$, and hence GW time-delay cosmography can extinguish this error term and set the measurement floor of $H_0$ to $\lesssim10$m/s/Mpc. Similar, the sub-second timing for gamma ray bursts (GRBs) and fast radio bursts (FRBs) provide a timing precision that can suppress the arrival time difference error term.  The sources of GRBs are believed to be binary compact object mergers or the collapse of massive stars, and the sources of FRBs are hypothized as magnetars. The detected in gamma rays or radio do not require a further detection in gravitational waves.

\subsection{Astrometric precision}
Beyond a measured time delay and an accurate lens model, there is an additional key requirement for time-delay cosmography, which is often either ignored or not sufficiently highlighted: the knowledge of the relative astrometry of the arriving images. To predict the Fermat potential differences (Eqn.~\ref{eq:fermat}) between two images, the position of where in the image plane ($\vec{\theta}$) the Fermat potential is evaluated, matters. 

Following \cite{Birrer:2019}, to first order, a change in the position of an image (A) in the image plane $\delta \vec{\theta_{\rm A}}$ leads to a change in the predicted relative Fermat potential between two images, A and B, $\Delta\phi_{\rm AB}$
\begin{equation}
    \delta \Delta\phi_{\rm AB}(\delta \vec{\theta_{\rm A}}) \approx \left(\vec{\theta}_{\rm B} - \vec{\theta}_{\rm A} \right) \frac{\partial \vec{\beta}}{\partial \vec{\theta}_{\rm A}}
\delta \vec{\theta}_{\rm A}.
\end{equation}
In regard to a relative error on the Hubble constant, the uncertainty in the astrometry propagates as

\newcommand{\arcsec}{$^{\prime\prime}$}
\begin{equation} \label{eqn:error_h0}
\frac{\sigma_{H_0}}{H_0} \approx
\frac{D_{\Delta t}}{c} \frac{\theta_{\rm AB}}{\Delta t_{\rm AB}} \frac{\delta \theta}{\mu},
\end{equation}
with $\mu$ being the absolute lensing magnification and $\theta_{\rm AB}$ is the image separation. For a typical redshift configuration, an image separation of 1\arcsec\ and a time delay of 1 day \cite{Wojtak:2019, Smith:2023}, the requirement on the relative astrometry to not cause more than a 5\% uncertainty on $H_0$ is about 0.6 mas \cite{Birrer:2019}.
We further refer to \cite{Birrer:2019} for details on the astrometric requirements for time-delay cosmography.

Sky localisation of dark sirens without an EM counterpart will be very poor and far beyond the astrometric requirement, and hence will not contain useful cosmographic information, even in the far future\footnote{The only potentially valuable path is for lensing configurations that are extremely simple (such as a single point mass lens) such that the image positions could be inferred through the time-delay measurements.}.
Cosmographic estimates with time delays require an EM counterpart, currently only observed for binary neutron star--neutron star (BNS) mergers \cite{NSNSGW:2017}. 
In the future, it may be possible to localise black hole-black hole (BBH) or black hole-neutron star (BH-NS) mergers if the putative association of some mergers with flares in quasar accretion disks is confirmed \cite{Graham:2020}.

Within this decade, the most suited instrument to discover EM counterparts of BNS mergers, whether lensed or not, is the Vera C. Rubin Observatory \cite{LSST19} due to its large collecting area and field of view, and fast slew speed. In addition to the Legacy Survey of Space and Time's (LSST's) wide fast deep survey of 18,000 square degrees, and the deep drilling fields, the observatory envisages spending up to 3\% of observing time on target of opportunity observations [\href{https://pstn-055.lsst.io/}{SCOC Phase 2 Recommendations}] motivated by several science goals including follow-up of GW sources whether lensed or not \cite{Margutti:2018, Smith:2019, Andreoni:2022, Smith:2023}. 
The astrometric precision of the Vera C. Rubin Observatory may provide few mas astrometric precision even for faint point sources. Details of how precise the measurements can be achieved remain unclear. For bright quasars observed with the \textit{Hubble} Space Telescope, a comparison between \textit{Gaia} astrometry and different modelling approaches result in an astrometric scatter of $\sim 1.7$ mas in each direction \cite{Ertl:2023}. Transient point sources might be more precise since a template without the transient source is available.
A $\lesssim 1$mas astrometry on a lensed kilonova (KN) from \emph{JWST} observations [Ryczanowski et al., this volume] (or lensed AGN accretion disk flare) implies that astrometric uncertainties on $H_0$ from GW time-delay cosmography will be no worse than for ``conventional'' time-delay cosmography for arrival time differences of order a day or longer.

As the rates of lensed GW are anticipated to be low, the exploration of larger cosmic volume where lensing is more likely (at higher redshift) will be necessary. The next generation ground-based GW detectors (Cosmic Explorer, Einstein Telescope) will be able to probe events close to the cosmic horizon. At these distances, optical EW counterparts are expected to be very faint. In this scenario, looking for GRB + GW pairs could be more productive as a route to find localizable lensed GW events.

Considering a time-delay measurement with a precision of 1 ms (see Section~\ref{sec:time-delay}), a displacement of an image of order $10^{-7}$ mas changes the time delay by the same amount.
Creative approaches to how one can interpret time delays of order 1 ms may be developed.
For example, when knowing the lens model, a ultra-precise time-delay measurement can, at least in principle, be used to constrain the relative image positions to unprecedented relative astrometric precision. Such astrometric precision might be able to constrain cosmological parallax (e.g. similar to \cite{Pierce:2019}).

%\subsection{Small-separation lenses}

\subsection{Small Einstein Radius lenses}

A significant fraction of the gravitational lensing optical depth resides behind galaxy scale lenses (e.g., \cite{Robertson:2020}), which are associated with typical Einstein radii of $\simeq0.5$\arcsec\ \cite{Collett:2015}. For time-delay cosmography with the classical EM sources, this regime is of lesser use, since the predicted time delays are short, and the limited ability to measure time delays below a day limits the precision from those systems.
This was underlined by the first galaxy-scale lensed supernova \cite{Goobar:2017,Goobar:2023}, discovered by a magnification method, hence not subject to an angular separation bias, has a very small separation of 0.3\arcsec\ and a predicted time delay is of order hours with the maximum delay below 35 hours at 99.9\% confidence. With the time-delay measurement not being the limiting factor for lensed GW, this gives access to using small Einstein radius lenses for time-delay cosmography, and potentially pushing to lower mass lenses, such as spirals or even dwarf galaxies.

\subsection{Lensing degeneracies}\label{subsection:degeneracies}

One of the key systematics uncertainties in time-delay cosmography is the mass-sheet degeneracy (MSD; \cite{Falco:1985}). This degeneracy is also present when using lensed GWs. Absolute magnification constraints for standardizable NS--NS star merger can similarly constrain the MSD as can standardizable lensed supernovae of Type Ia. Given the standardizable uncertainty of single NS--NS star mergers is of order $\sim 10\%$ \cite{NSNSGW:2017}, a significant number of standardizable lensed GW events are required, with the same caveats of (stellar) microlensing magnifications \cite{Foxley-Marrable:2018}.

In the geometrical optics scenario, the mass-sheet degeneracy cannot be broken with strong lensing-only information. 
Stellar kinematics measurement of the deflector (e.g., \cite{Treu:2002, Koopmans:2003, Birrer:2020, Shajib:2023}), standardisable magnifications with type Ia supernovae (e.g., \cite{Kolatt:1998, Oguri:2003, Foxley-Marrable:2018, Birrer:2021glSNe}) or galaxy-galaxy weak gravitational lensing measurements \cite{Khadka:2024} can break the mass-sheet degeneracy.  

For lensed GWs, \cite{Cremonese:2021} finds that in the interference regime, and, in part, in the wave-optics regime, the mass-sheet degeneracy can be broken with only one lensed waveform, thanks to the characteristic interference patterns of the signal that has a different response to the change in the source position and/or lens mass under the MSD. This is a promising regime for the first discovery of a gravitationally BNS/KN because the predicted arrival time distribution has significant power on timescales for which the lensed strain signals would be detectable by the GW detectors \cite{Smith:2023}, and thus signatures of interference could be detected.

\section{Lens modeling and numerical forecasts}\label{sec:forecast}
In Section~\ref{sec:gw}, we identified and individually discussed different aspects of time-delay cosmography with lensed GW signal, in particular the challenge with astrometry and the opportunity with high-precision time delays. Separating these different aspects allows us for a first-order qualitative assessment of the relative importance of different aspects.
Ultra-precise time delays can inform the lens model to some extent partially compensating for the lack of astrometric precision.
In this section, we perform a combined quantitative forecasts including lens model uncertainties in combination of different levels of precision in astrometry and time-delay measurement. 

\subsection{Model setup}
\begin{figure}[ht]
\centering
\includegraphics[width=0.5\textwidth]{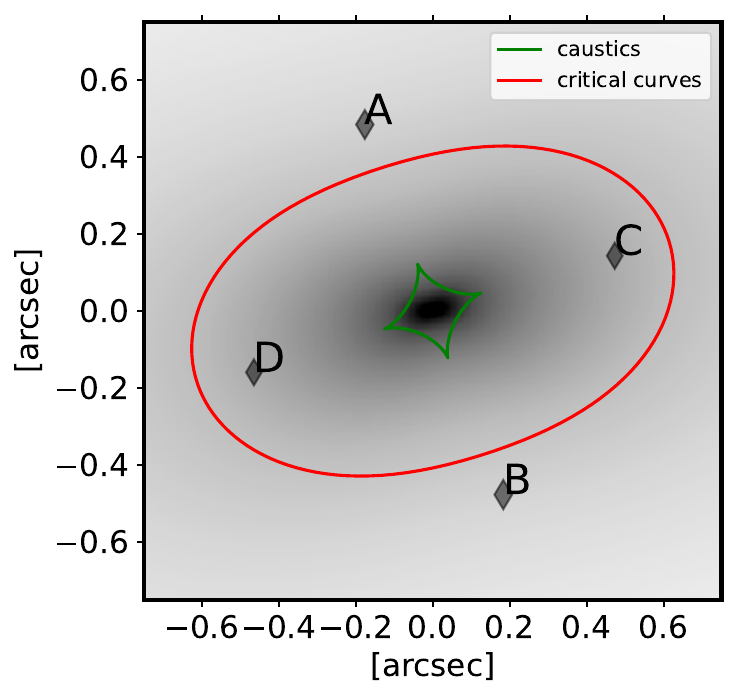}
\caption{Lensing configuration used for the forecast. The lens has an Einstein radius of 0.5''. The lensing configuration results in relative time delays of order $\sim 3$ days between the images.}
\label{fig:lens_example}
\end{figure}

We set up a quadruply imaged GW event with an Einstein radius of $0.5''$, the typical expected Einstein radius expected without any additional pre-selection being made on the discovery probability (e.g., \cite{Collett:2015}). Such an image configuration results in predicted time delays of order $\sim$ 2 days. While being among the most common configurations, such short delays do not provide sufficient constraints on $H_0$ with optically measured time delays. For the lens model, we use a elliptical power-law mass distribution with external shear, a common model being used.
For the priors, we generally chose priors that mimique constraints from imaging data of space-based optical telescopes, such as HST or JWST.
For the power-law slope we chose a Gaussian prior with an uncertainty of $\sigma_{\rm \gamma} = 0.1$. We assume that we can measure and deduce the centroid of the deflector model from an associated light component with an uncertainty of $0.01''$ in each angular direction. We assume a measurement of the Einstein radius to $0.01''$, for the ellipticity components $0.05$ and for the external shear components $0.01$ at one sigma. Overall, these priors/constraints allow for constraints of $H_0$ of $\sim 12\%$. This uncertainty does not including astrometric and time-delay uncertainties, nor information of the astrometry or time delays that can help constrain the lens model. Figure \ref{fig:lens_example} illustrates the lens configuration we use in our example forecast.

For the astrometric position of the arriving images of the time-variable event we use a set of different assumptions with one-sigma precision in the range of [1'', 0.1'', 5 mas, 0.1 mas]. The 1'' precision effectively ereases any meaningful constraints on the lens model or relative image position. A precision of 0.1'' corresponds on a detected faint optical signal from a ground-based telescope. 5 mas precision corresponds to the level of astrometry with lensed quasars and SNe and is at the level where substructure millilensing affects the image positions. 0.1 mas corresponds to a not yet achieved but possibly feasible astrometry for well-detected transient events in the optical.

For the time-delay measurement, we assume an array of different precisions [1 day, 1 hour, 1 min, 1 ms]. A one-day precision is about the limit of an optical monitoring campaign (\cite{Millon:2020}). A one-hour precision corresponds to about the fluctuations caused by substructure (\cite{Keeton:2009, Gilman:2020}).
A one-minute precision corresponds to roughly micro lensing of stars, and the one millisecond corresponds to the precision of the GW detectors.

We are using \textsc{lenstronomy}\cite{lenstronomy, lenstronomyII} to perform the posterior inference of the lens model, image positions, and time-delay distance (\ref{eq:ddt}) taking into account the lens model priors and astrometric and time-delay measurements.

\subsection{Results and implications}

To investigate the role of the uncertainty in the lens model, we perform two mock inferences, one with fixed lens model, and a second inference varying simultaneously the lens model in combination with the astrometric and time-delay data. 

\begin{figure}[ht]
\centering
\includegraphics[width=0.8\textwidth]{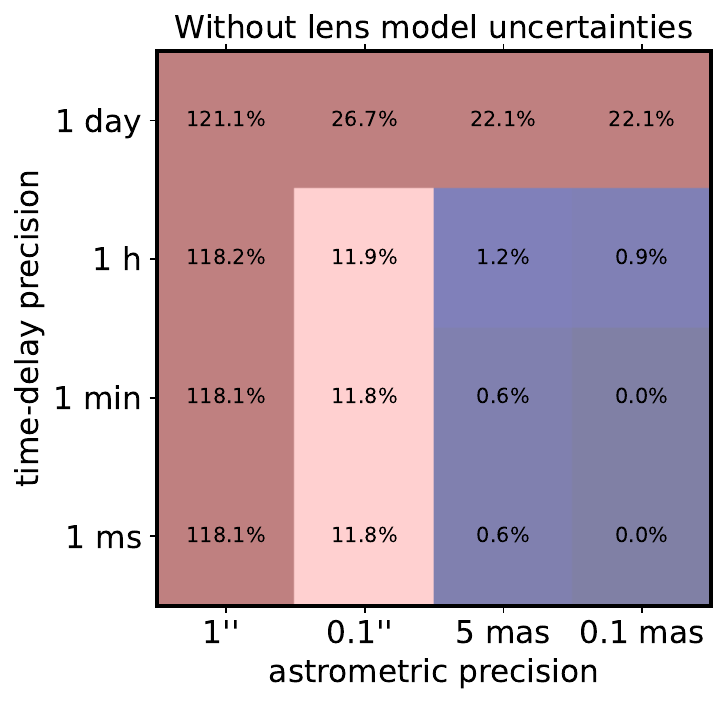}
\caption{Forecast on $H_0$ precision of a typical quadruply lensed configuration with an Einstein radius of 0.5'' as a function of astrometric precision in the lensed images and the time-delay measurement precision, only accounting for the astrometric and time-delay uncertainties, under perfect knowledge of the lens model.}
\label{fig:error_budget_fixed_lens}
\end{figure}

In the first case, we fix the lens model to the truth and let only the time-delay and astrometric uncertainties impact the precision on $H_0$. In that case, the uncertainties can be calculated by adding in quadrature the astrometric and the time-delay uncertainties for each image pair, with the astrometric uncertainty based on \cite{Birrer:2019}. Figure~\ref{fig:error_budget_fixed_lens} shows the expected uncertainties on $H_0$ for the grid of time-delay and astrometric precision for the scenario without lens model uncertainties. To achieve a sub 10\% uncertainty in $H_0$, the key is the astrometric requirement, while a time-delay measurement precision of $\sim$1 hour is sufficient to provide a $\sim$1\% error on $H_0$.

\begin{figure}[ht]
\centering
\includegraphics[width=0.8\textwidth]{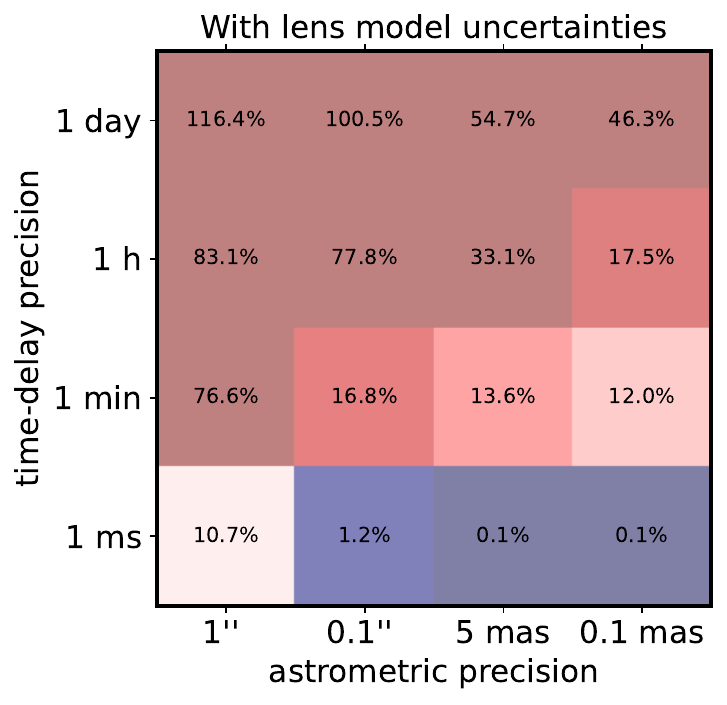}
\caption{Forecast on $H_0$ precision of a typical quadruply lensed configuration with an Einstein radius of 0.5'' as a function of astrometric precision in the lensed images and the time-delay measurement precision including lens model uncertainties. With ultra-precise time-delay measurements, the time delays can provide sufficient information on the astrometry and the lens model, effectively beating the noise floors of the astrometry and the lens model priors, simultaneously.}
\label{fig:error_budget}
\end{figure}

Figure~\ref{fig:error_budget_fixed_lens} only tells part of the story. Figure~\ref{fig:error_budget} presents the full joint posterior inference on $H_0$ from our simulated quadruply lensed transient with the lens model uncertainties according to our priors based on imaging modelling taking into account the full covariances. Precise time delays of order a millisecond can lead to $H_0$ uncertainties of order 10\% even in the case of the absence of precise astrometry. At first, this may sound surprising, as the achieved precision is even better than what is expected from the lens model priors on the Fermat potential. With ultra-precise time-delay measurements, the time delays can provide sufficient information on the astrometry and the lens model, effectively beating the noise floors of the astrometry and the lens model priors, simultaneously.

The result presented in Figure~\ref{fig:error_budget} holds for the model complexity of an elliptical power-law mass density with external shear. More complex lens configurations that require more degrees of freedom will result in ever more independence of the astrometric and the time-delay error budget.
On the other hand, the lens model priors chosen were relatively conservative. A hierarchical analyses and modelling of population of lenses (e.g.,\cite{Birrer:2020, Erickson:2024}) and the detailed modelling of the lens once the transient has faint away (e.g., \cite{Ding:2021}) can possibly improve the lens model constraints.
We also note a de-facto noise floor for the time-delay prediction caused by small mini-halos along the line of sight and subhalos in the main deflector. While this noise floor can diminish the return for time-delay cosmography, it can be turned into a powerful tool for dark matter substructure analyses (e.g., \cite{Keeton:2009, Gilman:2020}). We suggest further exploration in the direction substructure constraints in particular using lensed GW.

The numerical exercise stated in this section did not include the effect of the MSD (Section ~\ref{sec:gw}\ref{subsection:degeneracies}). The MSD and the associated radial density profile will likely remain the dominant source of uncertainty in time-delay cosmography.
At the same time, even with the presence of an MSD, with an ultra-precise time delay, one can probe relative distance ratios between two or more lenses, as the ratio of time-delay distances is invariant under a joint MSD effect between multiple lenses.

\section{Summary}\label{sec:summary}
In this manuscript we have highlighted key challenges and opportunities of using multiply-imaged lensed GW for time-delay cosmography.
We highlighted the similarities of using GW signals to be applied to time-delay cosmography. We identified the astrometric precision requirement of the images as a key challenge to overcome and emphasized the potentially significant impact that near-perfect time-delay measurements of lensed GW can bring to the table. Ultra-precise time-delay measurements of quadruply lensed systems can also provide information about the location (astrometry) of the arriving images and inform lens models, circumventing at least partially the astrometric requirements.
A potentially transformative probe of lensed GW could emerge when exploiting a completely different class of strong lenses: small-separation lenses, or lenses with a very small-separation image pairs, where wave-optics effects come into play and can be utilized.

\dataccess{The article has no additional data.}
\aiuse{We have not used AI-assisted technologies in creating this article.}
\aucontribute{S.B.: conceptualisation, formal analysis, funding acquisition, methodology, project administration, visualisation, writing---original draft; G.P.S.: writing--review and editing; A.J.S.: writing--review and editing; D.R.: writing--review and editing; N.A.:
\\
All authors gave ﬁnal approval for publication and agreed to be held accountable for the work performed therein.}
\conflict{We declare we have no competing interests.}
\funding{G.P.S. acknowledges support from the Leverhulme Trust and the Science and Technology Facilities Council. 
G.P.S acknowledges support from The Royal Society, the Leverhulme Trust, and the Science and Technology Facilities Council (grant number ST/X001296/1).
Support for this work was provided by NASA through the NASA Hubble Fellowship grant HST-HF2-51492 awarded to A.J.S. by the Space Telescope Science Institute, which is operated by the Association of Universities for Research in Astronomy, Inc., for NASA, under contract NAS5-26555.}
\ack{S.B., G.P.S., D.R., A.J.S. and N.A. thank The Royal Society for their support.}

%%%%%%%%%% Insert bibliography here %%%%%%%%%%%%%%
\section*{Affiliations}
$^{1}$Department of Physics and Astronomy, Stony Brook University, Stony Brook, NY 11794, USA\smallskip\\
$^{2}$School of Physics and Astronomy, University of Birmingham, Edgbaston, B15 2TT, UK\smallskip\\
$^{3}$Department of Astrophysics, University of Vienna, Türkenschanzstrasse 17, 1180 Vienna, Austria\smallskip\\
$^{4}$Department of Astronomy and Astrophysics, University of Chicago, Chicago, IL 60637, USA\smallskip\\
$^{5}$Kavli Institute for Cosmological Physics, University of Chicago, Chicago, IL 60637, USA\smallskip\\
$^{6}$NHFP Einstein Fellow\smallskip\\
$^{7}$Institute of Cosmology and Gravitation, University of Portsmouth, Burnaby Rd, Portsmouth, PO1 3FX, UK\smallskip\\
$^{8}$Oskar Klein Centre, Department of Physics, Stockholm University, SE-106 91 Stockholm, Sweden

\bibliography{references}  
\bibliographystyle{RS}

\end{document}